\documentclass[journal]{IEEEtran}
\usepackage{cite}

\ifCLASSINFOpdf
  \usepackage[pdftex]{graphicx}
\else
  \usepackage[dvips]{graphicx}
\fi
\usepackage{amsmath}
\usepackage{mathtools}
\usepackage{caption}
\usepackage{subcaption}
\usepackage[ruled,vlined]{algorithm2e}
\usepackage{comment}
\usepackage{algpseudocode}

\usepackage{multirow}
\usepackage{multicol}
\usepackage[table,xcdraw]{xcolor}

\usepackage{hyperref}
\usepackage{amssymb}
\usepackage{tikz}
\newcommand\copyrighttext{%
  \footnotesize This work has been submitted to the IEEE for possible publication. Copyright may be transferred without notice, after which this version may no longer be accessible}
\newcommand\copyrightnotice{%
\begin{tikzpicture}[remember picture,overlay]
\node[anchor=south,yshift=10pt] at (current page.south) {\fbox{\parbox{\dimexpr\textwidth-\fboxsep-\fboxrule\relax}{\copyrighttext}}};
\end{tikzpicture}%
}

\begin{document}
%
\title{Removing the need for ground truth UWB data collection: self-supervised ranging error correction using deep reinforcement learning}
%
%
%

\author{Dieter Coppens, Ben Van Herbruggen, Adnan Shahid, \IEEEmembership{Senior member , IEEE}, Eli De Poorter
\thanks{D. Coppens, B. Van Herbruggen, A. Shahid and E. De Poorter are with IDLab, Department of Information  Technology, Ghent University—imec, 9052 Ghent,
Belgium (e-mail: dieter.coppens@ugent.be;
adnan.shahid@ugent.be)}
}

%
%


\markboth{Journal of \LaTeX\ Class Files,~Vol.~14, No.~8, August~2015}%
{Shell \MakeLowercase{\textit{et al.}}: Bare Demo of IEEEtran.cls for IEEE Journals}
%



\maketitle
\copyrightnotice
\begin{abstract}
Indoor positioning using UWB technology has gained interest due to its centimeter-level accuracy potential. However, multipath effects and non-line-of-sight conditions cause ranging errors between anchors and tags. Existing approaches for mitigating these ranging errors rely on collecting large labeled datasets, making them impractical for real-world deployments. This paper proposes a novel self-supervised deep reinforcement learning approach that does not require labeled ground truth data. A reinforcement learning agent uses the channel impulse response as a state and predicts corrections to minimize the error between corrected and estimated ranges. The agent learns, self-supervised, by iteratively improving corrections that are generated by combining the predictability of trajectories with filtering and smoothening. Experiments on real-world UWB measurements demonstrate comparable performance to state-of-the-art supervised methods, overcoming data dependency and lack of generalizability limitations. This makes self-supervised deep reinforcement learning a promising solution for practical and scalable UWB-ranging error correction.
\end{abstract}


%
\IEEEpeerreviewmaketitle

\section{Introduction}
%
%
%
%
\IEEEPARstart{P}{recise} indoor positioning technology has attracted significant research interest in recent years due to its role in overcoming the limitations of global positioning system (GPS) in indoor environments for Internet of Things (IoT) applications such as assistive healthcare systems \cite{bazo2021survey}, sports tracking \cite{minne2019experimental}, smart logistics \cite{elsanhoury2022precision}  and various location-based services \cite{huang2018location}.  Following this trend, Ultra-Wideband (UWB) technology has seen a surge in interest and become one of the more promising technologies for indoor positioning systems (IPS). UWB IPS can achieve centimeter-level positioning accuracy due to the wide bandwidth ($>$500 MHz) and very short time duration of the pulse (around 2 ns) \cite{coppens2022overview}. While these signal characteristics make UWB more resilient to multipath effects (compared to traditional narrowband techniques such as SigFox, LoRa, Narrowband Internet of Things (NB-IoT), etc. \cite{alarifi2016ultra,wymeersch2012machine}). However, a major remaining challenge is correcting ranging errors caused by this multipath behavior in non-line-of-sight (NLOS) conditions \cite{gifford2020impact}, \cite{denis2003impact}. Current methods to detect and reduce errors caused by NLOS conditions rely mostly on machine/deep learning models trained using large datasets of UWB ranges and raw physical data like the channel impulse response (CIR) \cite{mao2018probabilistic,jaron,li2023variational} or calculated features \cite{wymeersch2012machine} (e.g. amplitude of the signal, energy, power ratio, etc.) labeled with the true positions. While these approaches can lead to high performance, it comes with two major disadvantages. First, collecting such labeled data requires a tedious labeling effort and dataset collection, which requires specialized equipment and personnel with expertise in UWB positioning and ground truth data collection. Second, the usability is limited by the generalization problem, the accuracy of trained solutions drops severely in unseen environments. The unseen environments have different anchor topologies, different sizes, and different UWB hardware, or contain different types of objects that degrade the performance due to (1) variations in the CIR and (2) different UWB physical layer (PHY) properties. The generalization problem worsens the data collection problem as each unique environment requires the labeling effort to be repeated and even so, the environment may have changed by then. These two disadvantages have previously been addressed using (1) semi-supervised learning \cite{jaron} \cite{li2021semi}, to reduce the data collection and (2) transfer learning to enable better performance in unseen environments while using only a few labeled samples \cite{fontaine2023transfer,li2023unsupervised}. However, all these approaches still require some labeled samples, thus a tedious data collection effort. One other research proposes a self-supervised ranging error correction \cite{yangselfsupervised} that does not require ground truth or label collection. It uses classical location approaches to estimate the location and range with a deep network jointly. However, the learning here is limited as they do not use signal features to aid and improve the learning process, and they cannot correct separate ranges. 
\begin{figure}[ht]
    \centering            \includegraphics[width=0.84\linewidth]{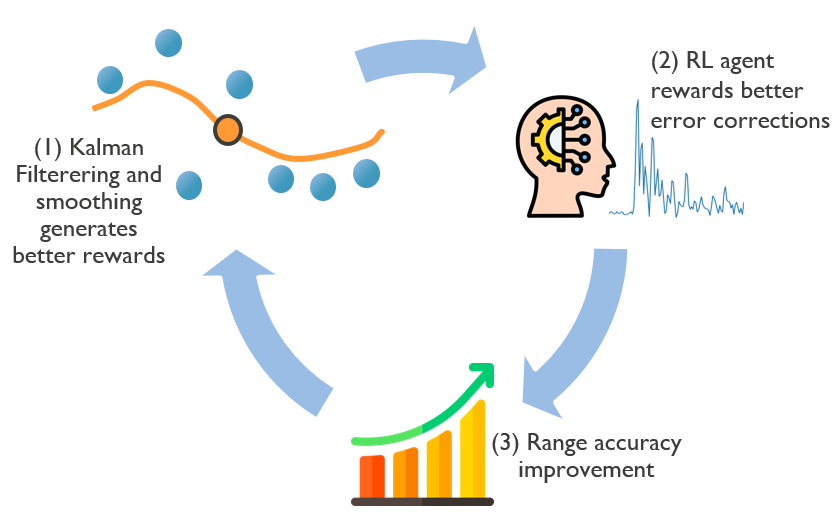}
    \caption{Conceptual illustration of the idea behind UWB ranging error correction}
    \label{fig:overview}
\end{figure}

To address these shortcomings, we propose a novel approach based on deep reinforcement learning (RL) which relies on using iteratively improving information automatically derived, removing the need for exact labels. We assume occasional movements of people or vehicles in the environment, which follow sufficiently predictable trajectories. By combining this predictability with filtering, smoothing, and error correction, improvements in error correction are rewarded over time. This iterative process continually enhances the filtered and corrected positions, leading to continuously improving the available information for ranging correction. This concept is illustrated in Figure \ref{fig:overview}. Finally, the filtering and smoothing can be removed to provide real-time range error correction.

The main contributions of this paper are:
\begin{itemize}
    \item Introduction of the first self-supervised deep RL framework for CIR-based UWB ranging error correction in a two-way ranging (TWR) system.
    \item The self-supervised nature of this framework eliminates the requirement for data collection or reliance on ground truth for successful implementation.
    \item Analyzing the performance of our self-supervised deep RL framework compared to a state-of-the-art supervised convolutional neural network (CNN)
\end{itemize}
The remainder of the paper is organized as follows. Section \ref{sec:relatedwork} discusses the related work for UWB range error correction. In Section \ref{sec:dataset}, the environment in which the dataset is gathered and how the measurements are performed is described. Next, in Section \ref{sec:problem} the UWB ranging error system model and problem are described. Section \ref{sec:methodology} describes the proposed RL methodology and Section \ref{sec:results} discusses the performance of the proposed algorithms. The future work follows this in Section \ref{sec:future} and finally the conclusion in Section \ref{sec:conclusion}.

\section{Related work}
\label{sec:relatedwork}
In this section, an overview of related papers for UWB range error correction in the literature is provided. The related work is split up into four categories, (1) supervised learning, (2) semi-supervised learning, (3) transfer learning, and (4) self-supervised learning.
\subsection{Supervised learning}
The authors of \cite{wymeersch2012machine} propose a feature-based approach with both support vector machine (SVM) regression and a Gaussian process (GP) to form an estimate of the ranging error.
The authors of \cite{li2023variational} propose an approach using latent variables that encapsulate information from the CIR about both distance and environmental features to then employ variational inference techniques with neural networks to perform approximate inference in a supervised manner.
The authors of \cite{mao2018probabilistic} propose a supervised deep learning approach for UWB-ranging error correction. It leverages a probabilistic deep learning architecture by combining variational inference with probabilistic neural networks. The approach uses a variational autoencoder to learn features from the CIR. 
\cite{jaron} uses a similar autoencoder approach for feature extraction from the CIR. Still, the models are trained in a dual-loss fashion to optimize unsupervised autoencoding and supervised prediction jointly.
While both \cite{jaron} and \cite{mao2018probabilistic} leverage unsupervised pre-training of the autoencoder layers, the key ranging error prediction task is formulated as a supervised learning problem. Here, labeled data is used to train a model to directly map inputs to known target outputs. These papers show that supervised machine learning approaches using both raw physical data (CIR) or features can be used to significantly improve the UWB ranging performance. However, none of them address the problem of data collection or the generalization problem, meaning that real-world usability is limited.
\subsection{Semi-supervised learning}
The authors of \cite{li2021semi} propose a semi-supervised approach for UWB-ranging error mitigation. Similar to \cite{li2023variational} it formulates the problem with a latent variable that encapsulates information about both ranging error and environment. It utilizes a loss function composed of supervised and unsupervised terms, meaning it can use information from both labeled and unlabeled data.
This paper addressed the data collection problem and partly succeeded by using semi-supervised learning, meaning that less labeled data is necessary. However, it is not complete unsupervised learning and still requires some supervising (data labeling).

\begin{table*}[]
\centering
\caption{Comparison of our proposed UWB ranging error correction approach with related work. The table mentions the learning method and inputs for learning that are used.}
\label{tab:related work}
\begin{tabular}{c|ccccccc}
\textbf{Paper} &  \begin{tabular}[c]{@{}c@{}}\textbf{Self-} \\ \textbf{supervised}\end{tabular}  & \textbf{ML approach} & \begin{tabular}[c]{@{}c@{}}\textbf{Localization} \\ \textbf{technique}\end{tabular} & \textbf{Environment type} & \textbf{\#Anchors} & \textbf{Input} &  \begin{tabular}[c]{@{}c@{}}\textbf{Output} \\ \textbf{of model}\end{tabular}  \\ \hline
\cite{wymeersch2012machine}& &\begin{tabular}[c]{@{}c@{}}SVM \\ Gaussian process\end{tabular} &TWR&LOS/NLOS&5 &Features, range&Ranging error\\
\cite{li2023variational}& &\begin{tabular}[c]{@{}c@{}}Inter-Instance \\ Variational Auto-Encoder\end{tabular}&TWR&\begin{tabular}[c]{@{}c@{}}LOS/NLOS \\ (Simulated)\end{tabular}&10 &CIR&\begin{tabular}[c]{@{}c@{}}Ranging Error \\ Environment label\end{tabular}\\
\cite{mao2018probabilistic}& &\begin{tabular}[c]{@{}c@{}}Variational inference \\ Probabilistic learning\end{tabular} &TWR&LOS/NLOS&1 &CIR&Ranging error\\
\cite{jaron}& & \begin{tabular}[c]{@{}c@{}}Variational \\ Auto-Encoder\end{tabular}&TWR&LOS/NLOS&19 &CIR & Ranging error\\
\cite{li2021semi}& &\begin{tabular}[c]{@{}c@{}}Variational \\ Bayesian process\end{tabular}  & TWR & LOS/NLOS& 4&CIR&\begin{tabular}[c]{@{}c@{}}Ranging Error \\ Environment label\end{tabular}\\
\cite{fontaine2023transfer}& &Transfer learning&  TWR &LOS/NLOS&21 &Features, CIR& \begin{tabular}[c]{@{}c@{}}Ranging Error \\ LOS/NLOS label\end{tabular}\\
\cite{li2023unsupervised} & &\begin{tabular}[c]{@{}c@{}}Transfer learning\\ Domain Adversarial Training\end{tabular} & TWR & LOS/NLOS&1 & CIR &\begin{tabular}[c]{@{}c@{}}Ranging Error \\ Environment label\end{tabular}\\
\cite{yangselfsupervised}&\checkmark & CNN  & TOA & LOS& 4& Range &\begin{tabular}[c]{@{}c@{}}Ranging error \\ Positioning error\end{tabular}\\
\textbf{Our work} & \textbf{\checkmark} & \textbf{RL}& \textbf{TWR} & \textbf{LOS/NLOS}& 23 &\textbf{CIR}&\textbf{Ranging Error}
\end{tabular}
\end{table*}

\subsection{Transfer learning}
To address the generalization problem, the authors of \cite{fontaine2023transfer} propose a transfer learning (TL) framework for UWB error correction using feature- and raw CIR-based approaches. The framework allows for automatic optimizations for TL deep learning models towards new environments while keeping the number of labeled training samples small. The authors demonstrated high accuracy improvements (643 mm to 245 mm) with minimal data collection in challenging environments.
The authors of \cite{li2023unsupervised} propose an unsupervised TL method based on domain adversarial training and adaptive encoder-decoders. Domain adversarial training is applied to reduce the distribution mismatch between source and target environments. The method still requires labeled data for training the source model.
Transfer learning addresses the generalization problem, but still requires data collection for the pre-trained model and/or (minimal) data collection for transferring the knowledge to a new environment.
\subsection{Self-supervised learning}

To the best of our knowledge, \cite{yangselfsupervised} is the only self-supervised approach for UWB error mitigation. While both our method and \cite{yangselfsupervised} aim to improve ranging accuracy, there are significant differences in our approaches, driven by the distinct environments we target. The authors of \cite{yangselfsupervised} designed their method for simple, line-of-sight (LOS) environments with few anchors, while our approach addresses complex, large-scale industrial environments with both LOS and NLOS conditions. These environmental differences lead to contrasting algorithm designs. The authors of \cite{yangselfsupervised} propose a deep location and ranging correction (DLRC) network that jointly estimates tag position corrections and distance corrections using Time of Arrival (ToA). Their method requires 10 ranges for each anchor for a single error correction. This thus assumes the constant availability of all anchors with a high update rate, otherwise there is no correction available. In contrast, our method uses only one CIR and range per correction, allowing variable anchor availability and update rates typical in large-scale realistic environments. For instance, in our industrial testbed with 23 anchors, \cite{yangselfsupervised}'s approach would necessitate determining 230 ranges, leading to significant delays. Our method ensures faster processing by gathering only one range and its corresponding CIR. The information used for learning also differs significantly as the authors of \cite{yangselfsupervised} assume minimal tag movement between ranges, limiting the diversity of environmental effects captured. Our use of CIR as input encapsulates detailed information about the transmission channel, allowing our method to correlate the channel state with correction factors more effectively. While both approaches use similar principles in their loss/reward functions, there are notable differences. The method of \cite{yangselfsupervised} uses a dual loss function to simultaneously minimize position and range corrections. Our reinforcement learning-based method maximizes cumulative rewards based on the agreement between corrected ranges and Kalman filter positioning. This approach offers continuous adaptation capabilities, allowing our algorithm to adjust to environmental changes in real time – a crucial advantage in dynamic industrial environments.

\section{Dataset description}
\label{sec:dataset}

\begin{figure}[h]
\centering
 \begin{subfigure}{0.45\textwidth}
     \includegraphics[width=1\linewidth]{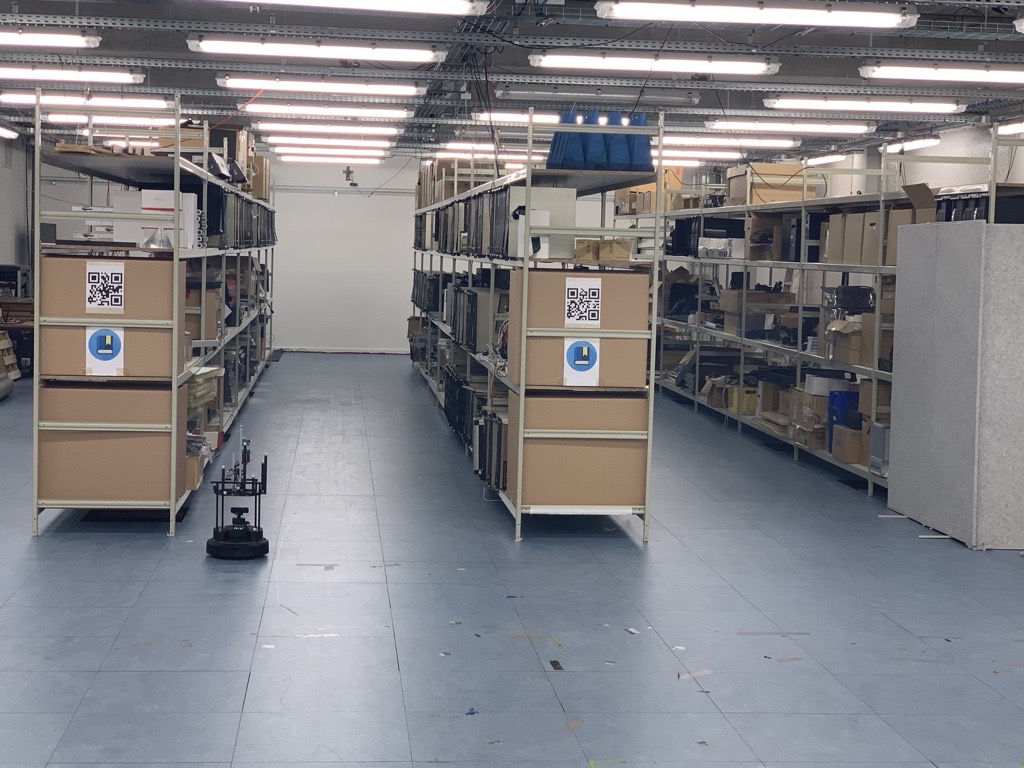}
     \caption{The IIoT lab environment}
     \label{fig:iiotpic}
 \end{subfigure}
 \begin{subfigure}{0.49\textwidth}
     \includegraphics[width=1\linewidth]{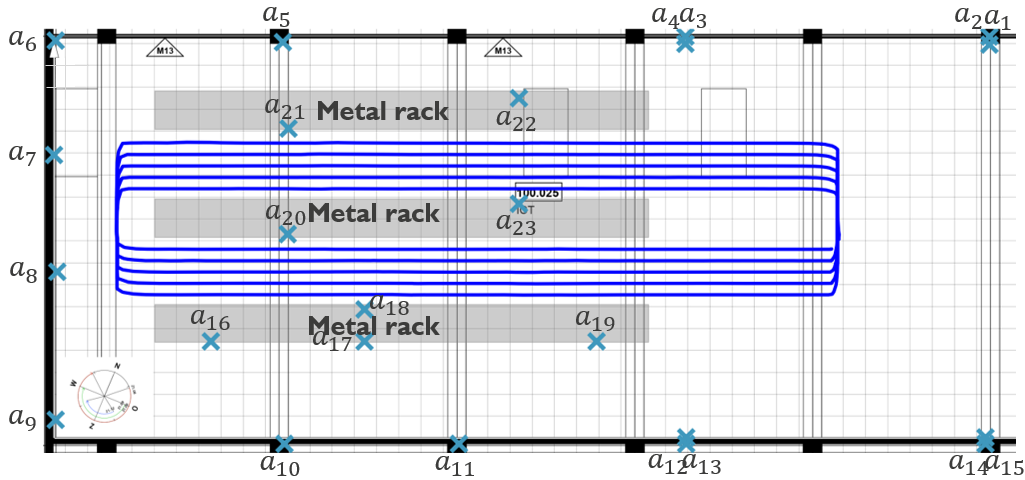}
     \caption{Floorplan of the IIoT lab with the position of each anchor indicated as light blue $X$ and the ground truth trajectory of the dataset as a dark blue line.}
     \label{fig:iiotfull}
 \end{subfigure}
\end{figure}

The dataset is collected in an industrial lab environment, which is part of the Industrial Internet of Things (IIoT) testbed \cite{iiot} of the IDLab research group at Ghent University. The lab is a 240 m\textsuperscript{2} warehouse environment, representative of many Industry 4.0 use cases. The IIoT testbed consists of an open space area and an area with metal racks, leading to LOS and NLOS situations, pictured in Figure \ref{fig:iiotpic}. The environment is equipped with 18 Qualisys Miqus M3 Motion Capture (MOCAP) cameras, capable of tracking hundreds of passive infrared reflective MOCAP markers with a quantified uncertainty in the millimeter range at speeds up to 340 Hz, enabling accurate ground truth determination for evaluation purposes (not used for training in this research). In addition, the MOCAP system is used in combination with a mobile robotic platform to drive repeatable trajectories through the lab. A total of 23 anchors are distributed over the environment, the placement is illustrated in Figure \ref{fig:iiotfull} with the light blue crosses. The dataset was collected using Wi-PoS devices \cite{wipos} that carry the Qorvo DW1000 UWB transceivers. During measurement, the CIR information used for learning was captured at the anchor nodes. To capture the data a mobile robot drives around the lab at 0.1 m/s, the trajectory of the robot is shown in Figure \ref{fig:iiotfull}. This trajectory leads to 3463 UWB ranging samples with the different anchors. The ranging method used in the system is called Asymmetric double-sided TWR (ADS-TWR) \cite{ADSTWR}
The same environment can change over time. To evaluate the performance of our proposed algorithm when there are changes in the environment, a second dataset was collected in the same warehouse 6 months later. At that time, there were more goods in the racks, additional clutter in the warehouse (obstacles, boxes, …) and the anchor nodes experienced many small disturbances over time. These combined effects lead to a more challenging environment. This second dataset is smaller and contains 1434 samples.

\subsection{Data pre-processing}
Before we use the CIR as state information in the RL algorithm, proposed in section \ref{sec:methodology}, we process the raw CIR data, in a pre-processing phase. The pre-processing of the raw CIR involves three distinct steps. First, the complex-valued IQ-sampled array is converted to an RSSI-sampled array. The RSSI is the absolute value of the complex IQ sample, by representing the real (I) and imaginary (Q) components on a Cartesian coordinate system, the RSSI value can thus be determined using the Euclidean distance from the origin:
\begin{equation}
    RSSI=\sqrt{I^2+Q^2}
\end{equation}
Second, the RSSI-array is trimmed to 150 samples, 50 samples before and 100 after the estimated first path by the DW1000 using the leading edge algorithm. Lastly, the remaining array is normalized using min-max normalization. We subtract the minimum value from each element to make the lowest value zero, and divide by the difference between the maximum and minimum values to scale the highest value to one:
\begin{equation}
    CIR_{norm} = \frac{CIR-min(CIR)}{max(CIR)-min(CIR)} 
\end{equation}
The normalization step results in smaller numerical values, which is better for training the RL algorithm because it improves the generalization capabilities. This approach tries to make the algorithm to learn to focus on learning signal-to-noise ratio (SNR) and peak features of the CIR, instead of absolute signal strength features. This is important as these can vary significantly across different settings and environments (for example, the average distance between tag and anchor in the environment or higher transmit power configurations) and may not necessarily indicate larger errors or (N)LOS signal propagation. The pre-processing steps, significantly reduce the complexity of the required models, making it computationally more efficient and faster to train. Additionally, a more focused input can help the model generalize better to new, unseen data, as it emphasizes learning essential features and patterns.
While reducing the input size, we focused the data around the first path where most errors occur \cite{fontaine2023transfer}.
\section{Problem and System description}
\label{sec:problem}
\begin{figure*}[]
    \centering
    \includegraphics[width=0.7\linewidth]{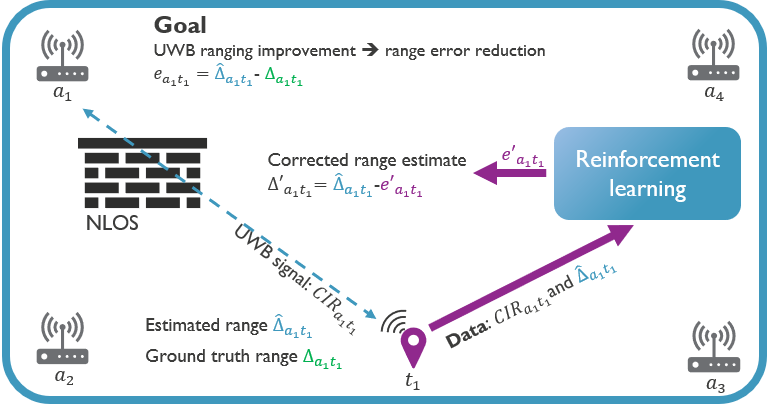}
    \caption{Illustration of the mathematical UWB localization system description}
    \label{fig:conceptoverview}
\end{figure*}

\label{sec:systemandproblem}
In this paper, the purpose is to correct the ranging measurements between the tag and anchor. For a better understanding of the problem, we first provide an overview of the UWB system.
\subsection{UWB Localization System}
An UWB IPS provides 3D positions $(x,y,z)$, relative to a reference point $ref=(0,0,0)$, for a tag $t_l \in \{t_1,t_2,...t_L\}$, with L the total number of tags. For this, it needs to know the coordinates of the fixed anchors $a_k \in \{a_1,a_2,...a_K\}$, with K the total number of anchors. To determine its position $t_{l_p} = (t_{l_x},t_{l_y},t_{l_z})$, the tag $t_l$ will measure the range (distance) to available anchors $a_k$ in the system. The ground truth range $\Delta a_k t_l$ can be expressed as follows:

\begin{equation}
    \Delta a_k t_l = \sqrt{(a_{k_x}-t_{l_x})^2+(a_{k_y}-t_{l_y})^2+(a_{k_z}-t_{l_z})^2}
\end{equation}
To find the position of the tag, $\Delta a_k t_l$ is estimated using time of flight (ToF). In this paper, ADS-TWR is used to estimate the ToF accurately. The ToF can be converted to  $\Delta a_k t_l$ as follows:
\begin{equation}
    \Delta a_k t_l = ToF_{a_k t_l} \cdot c
\end{equation}
Where c is the speed of light (3 x $10^8$ m/s). The ToF is typically estimated using the CIR which quantifies how the communication channel alters the UWB pulse, encapsulating its delay, amplitude, and phase changes. Using a leading-edge algorithm, as used in the popular DW1000 UWB chip \cite{DW1000}, the time when the arriving signal, from the accumulated UWB pulses, first rises above the noise floor is the detected first path ($fp'$). When there is no obstacle between anchor and tag, so-called LOS conditions, this detected first path is close to the actual first path ($fp$) and the ToF estimation is accurate. However, in real-world conditions, there are multipath effects and NLOS conditions. These two effects degrade first path detection and thus higher inaccuracies in ToF estimation. This effect can be demonstrated as follows using the CIR, logged at the UWB transceiver, for signal propagation between $a_k$ and $t_l$:
\begin{equation}
    CIR_{a_k t_l}(t) = \sum_{s=1}^{S}\alpha_s\delta(t-\tau_s) + n(t)
\end{equation}
Where t represents the timestamp for each value within the CIR (one CIR has 1016 complex values, corresponding to $\pm 10^-9 s$ each); S is the number of multipath components; $\alpha_s$ is the amplitude of the s-th multipath component; $\tau_s$ the time delay of the s-th multipath component; $\delta$ the Dirac delta function and $n(t)$ represents the additive white Gaussian noise (AWGN) present in the channel.
In NLOS conditions, $fp$ can be severely attenuated and the calculated ToF becomes inaccurate:
\begin{equation}
    \widehat{ToF}_{a_k t_l} = ToF_{a_k t_l} + \tau_{(fp'-fp)}
\end{equation}
With $fp'$ the first detected path above the noise floor and earlier multipath or the true first path are not detected. $\tau_{fp'-fp}$ is the time difference between the detected first path and the real first path. The ranging error this causes can be calculated as:
\begin{equation}
    e_{a_k t_l} = \tau_{(fp'-fp)} \cdot c
\end{equation}
The calculated range becomes:
\begin{equation}
    \widehat{\Delta a_k t_l} = \Delta a_k t_l + e_{a_k t_l}
\end{equation}
The goal of the UWB error correction model is to predict $e_{a_k t_l}$ as accurately as possible, using the CIR as input, without collecting a labeled dataset for the training process.
Because at each time step, there is only one range received and, for simplicity, in the remainder of the paper $a_k$ and $t_l$ will be omitted.
\section{Proposed methodology}
\label{sec:methodology}
\begin{table}
    \centering
    \caption{Mathematical symbols used throughout this article}
    \begin{tabular}{rl}
        \hline
       \textbf{Symbol}  & \textbf{Description} \\ \hline
        $ref$ & Localization system reference point \\
        $t_l$& UWB tag \\
        $a_k$ & UWB anchor \\
        L & Total number of tags\\
        K & Total number of anchors \\
        $\Delta a_k t_l$ & Euclidean distance between $a_k$ and $t_l$, shortened to $\Delta$  \\
        $ToF_{a_k t_l}$ &  Time of flight between $a_k$ and $t_l$, shortened to $ToF$\\
        $CIR_{a_k t_l}$ &  Channel impulse response, shortened to $CIR$\\
        $\widehat{Tof}_{a_k t_l}$ & Estimated ToF between $a_k$ and $t_l$, shortened to $\widehat{Tof}$  \\
        $e_{a_k t_l}$ & Ranging error between $a_k$ and $t_l$, shortened to $e$ \\
        $\widehat{\Delta a_k t_l}$ & Estimated range between $a_k$ and $t_l$, shortened to $\widehat{\Delta}$ \\
        $\widehat{e}$ & Estimated range error by RL agent \\
        $\Delta'$ & Corrected estimated range by RL agent \\
        $S_t$ & The state of the environment at time $t$ ($CIR$)\\
        $A_t$ & The action of the agent at time $t$ ($\widehat{e}$)\\
        $\pi$ & The agent's policy, $\pi: CIR \rightarrow \widehat{e}$ \\
        $\mu$ & The actor network of the RL agent \\
        $\theta$ & Weights of the actor network $\mu$ \\
        $Q$ & The critic network of the RL agent \\
        $\phi$ & Weights of the critic network $Q$ \\
        $y$ & Target Q-value \\
        $R_t$ & Reward received at time $t$ \\
        $\gamma$ & Discount factor, determining the weight of target critic\\
        $\dot{Q}$ & The target critic network \\
        $\dot{\phi}$ & Weights of the target critic network \\
        $\tau_{critic}$ & Soft copy factor of the critic \\
        $J$ & The sampled policy gradient \\
        $B$ & Number of samples in a batch \\
        $\epsilon$ & Exploration rate of the RL agent \\
        $\lambda$ & Decay factor of the exploration rate \\
        $\dot{\mu}$ & The target actor \\
        $\dot{\theta}$ & Weights of the target actor \\
        $\tau_{actor}$ & Soft copy factor of the actor \\
        $p_{EKF}$ & Extended Kalman Filter position \\
        $m$ & Middle position in smoothing buffer \\
        $N$ & Length of circular smoothing buffer\\ 
        $p_{avg,m}$ & Averaged position related to middle data $m$ in buffer \\
        $\Delta_{avg,m}$ & Resulting range from filtering and smoothing \\
        \hline
    \end{tabular}
    \label{tab:symbols}
\end{table}
In our methodology, we assume occasional movements of people or vehicles in the environment, which follow sufficiently predictable trajectories. Combining this predictability with filtering, smoothing, and error correction, improvements in error correction are rewarded over time. This is achieved using an RL process that continually enhances the filtered and corrected positions, leading to continuously improving data available for ranging correction.

\subsection{Reinforcement learning}
\label{sec:rl}
A RL framework consists of an agent and an environment interacting with each other. Anything in the area around the anchor and tag UWB devices that could affect range estimation is regarded as the environment. At each time $t$, the agent observes a state $S_t$ that represents all relevant available information of the environment and takes action $A_t$. Here, at each time step, the UWB localization system estimates the range between a tag and an anchor, $\widehat{\Delta_t}$. The RL agents corrects the estimate to $\Delta'_t = \widehat{\Delta_t } - \widehat{e_t}$, meaning that $A_t$ is the error correction $\widehat{e_t}$. $S_t$ is the $CIR_{t}$ associated with the current range estimation $\widehat{\Delta_t}$. We assume the UWB ranging error to be between $\pm1000$ mm, meaning that the action space $\mathcal{A}$ can be expressed as:
\begin{equation}
    \mathcal{A} = [-1000, 1000]
\end{equation}
The CIR value received from the DW1000 is pre-processed (described in detail in Section~\ref{sec:dataset}) to an array of 150 samples with a value between 0 and 1, meaning the state space $\mathcal{S}$ can be described as:
\begin{equation}
    \mathcal{S} = [0, 1]^{150}
\end{equation}
Due to the continuous action space, we base our custom RL algorithm on the deep deterministic policy gradient (DDPG) algorithm. The behavior of the agent is determined by the policy $\pi : \mathcal{S} \rightarrow \mathcal{A}$. The goal of reinforcement learning is to learn a policy that maximizes the expected rewards. DDPG uses an actor-critic framework, where the policy is determined by the actor network $\mu(S_t\mid\theta)$, with $\theta$ the weights of the network. The actor network approximates the optimal policy by learning to output the action that maximizes the expected cumulative reward. This expected cumulative reward is determined by an action-value function $Q(S_t,A_t)$ that is approximated by the critic network $Q(S_t,A_t\mid\phi)$, with $\phi$ the weights of the network, which takes in state-action pairs and estimates the Q-value. This critic is trained by minimizing the temporal difference between the predicted Q-value and the observed Q-value based on the received rewards $R_t$. This is done by minimizing the following loss function, which is an adapted version of the standard DDPG algorithm, as in this problem $A_t$ does not influence $S_{t+1}$:
\begin{equation}
    Loss(\phi) = (y_t - Q(S_{t},A_{t}\mid\phi))^2
\end{equation}
With $y_t$ the target Q-value
\begin{equation}
    y_t = R_t + \gamma\dot{Q}(S_{t},\mu(S_t\mid\theta)\mid \dot{\phi})
\end{equation}
This $y_t$ is dependent on target critic network $\dot{Q}(S_t,A_t\mid\dot{\phi})$, which is a slowly updated version of the main critic by softly copying the weights: $\dot{\phi} \leftarrow \tau_{critic} \phi + (1-\tau_{critic})\dot{\phi}$ with $\tau_{critic}\ll1$. This helps stabilize training in DDPG by providing a more consistent target for Q-value predictions. Not using a target critic can lead to increased sensitivity to non-stationary rewards and difficulties in achieving convergence. This would be catastrophic in this research, as the iterative update process causes non-stationary rewards. 
The actor and critic networks learn collaboratively: the actor network learns to maximize the predicted Q-values by the critic, simultaneously the critic network guides this learning by providing feedback on the quality of the chosen actions in corresponding states.
The actor network is updated using a sampled policy gradient to maximize the received expected cumulative rewards:
\begin{equation}
    J = -\frac{1}{B} \sum Q(S_t,A_t\mid\phi)
\end{equation}
With B the total number of samples in a batch.

\subsection{Action selection}
\begin{figure*}[]
    \centering
    \includegraphics[width=0.88\linewidth]{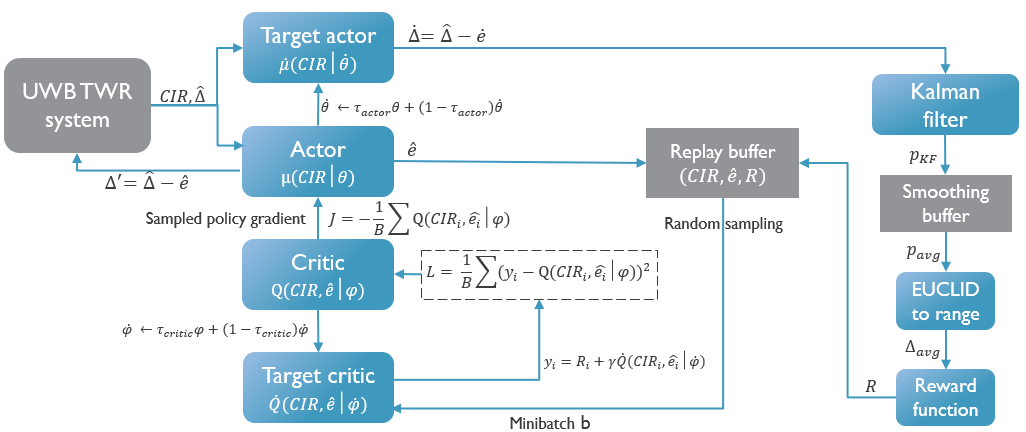}
    \caption{Complete overview of the proposed (adapted) DDPG algorithm for UWB error correction}
    \label{fig:completeoverview}
\end{figure*}
At each time step, the agent uses the actor network $\mu(S_t~\mid~\theta)$, to determine the current best estimate of the correction $\widehat{e_t}$ that will result in the highest reward. However, at the start, the actor network is not well-trained and does not yet know which actions will lead to the best rewards. This leads to two adaptations. First, the actor uses an exploitation/exploration step with an epsilon-greedy policy. In this policy, the correction from $\mu(S_t\mid\theta)$ is selected with a probability of 1-$\epsilon$ (exploitation of the actor). With a probability of $\epsilon$ a random action is chosen uniformly (exploration). The $\epsilon$ follows an exponential decay during training, at each step:
\begin{equation}
    \epsilon = \epsilon_{min} + (\epsilon_{max} - \epsilon_{min}) \cdot e^{-\lambda\cdot step}
    \label{eq:expdecay}
\end{equation}
With $\epsilon_{min}$ and $\epsilon_{max}$ the respective minimum and maximum exploration and $\lambda$ the decay. Exploration is a crucial aspect in RL because it allows the agent to explore which actions lead to good rewards without being constrained by what already has been learned. 
Second, to avoid bad training data for iterative improvement, we introduce a target actor, $\dot{\mu} (S_t\mid\dot{\theta})$ with weights initialized to zero, meaning that the first $\dot{e_t}$ will also be zero and will not influence the training data. Initializing the weights of a neural network to zero is generally avoided because it leads to a lack of symmetry breaking during training. When all weights are initialized to the same value, neurons in the network will have the same gradients during backpropagation, and they will continue to update in the same way. As a result, the network will fail to learn meaningful representations. However, the $\dot{\mu}(S_t,\dot{\theta})$ is not intended to be trained on, the weights from $\mu(S_t,\theta)$ will be "softly" copied to the target network once $\mu(S_t,\theta)$ has been trained sufficiently to improve the labels rather than deteriorate them. The soft target updates are given by: 
\begin{equation}
\dot{\theta} \leftarrow \tau_{actor}\theta + (1-\tau_{actor})\dot{\theta}
\end{equation} 
With $\tau_{actor}\ll1$. This poses the question of how to define "sufficiently trained". To address this, we employ the "ReduceLROnPlateau" scheduler from PyTorch \cite{pytorch}, a dynamic learning rate adjustment mechanism. The scheduler monitors the loss of the actor, reflecting the quality of the actor's policy, and adjusts the learning rate when a plateau in learning is detected. Once a plateau is identified, indicating that the actor network $\mu(S_t,\theta)$ has reached a state of sufficient training, the learning rate is reduced. This reduction triggers the soft updating of the target actor network $\dot{\mu}(S_t,\dot{\theta})$.

The soft target updates ensure a gradual and controlled transfer of knowledge from the actor network to the target actor network. It is crucial to note that during this soft updating process, the target actor $\dot{\mu}(S_t,\dot{\theta})$ does not participate in training the actions taken by the actor. Its role is confined to contributing to the data processing pipeline that leads to the calculation of rewards, and maintaining stability in the training process as illustrated in Figure \ref{fig:completeoverview}. In this way, the dynamic adjustment of the learning rate via "ReduceLROnPlateau" serves as a reliable criterion for defining "sufficiently trained" and triggers the appropriate updates to the target actor network.
\subsection{Data processing for self-supervised reward}
As discussed before, the actor determines ranging correction $\widehat{e_t}$ and the target actor generates correction $\dot{e_t}$ that leads to the corrected target range $\dot{\Delta}_t = \widehat{\Delta_t} - \dot{e_t}$ and this corrected range estimate is used to iteratively improve the range correction and self-generate better labels. $\dot{\Delta}_t$ is converted to a position using an Extended Kalman Filter (EKF)\cite{kalmanfilter}. 
\begin{equation}
p_{EKF,t} = \text{EKF}(\dot{\Delta}_t)    
\end{equation}
 Then added to a circular buffer $C$ of length N (assumed odd) used for smoothing. The size of the buffer impacts the smoothening and eventually the corrections, it is a trade-off, larger buffer size increases accuracy on straight paths but decreases accuracy in corners and other less predictable movements. If the buffer is full when a new $p_{KF,t}$ is added to the buffer together with its associated $\Delta'_t$, $CIR_t$ and $\widehat{e_t}$, the oldest value in the buffer is removed and the average position of all positions in the buffer is determined and linked to the value at the middle position of the buffer:
\begin{equation}
    m=t+\frac{N-1}{2}
\end{equation}
And average position:
\begin{equation}
    p_{avg,m} = (\frac{1}{N} \sum_{i=t}^{t+N-1} x_i, \frac{1}{N} \sum_{i=t}^{t+N-1} y_i)
\end{equation}
This $p_{avg,m}$ is related to the remaining data at position $m$ in the circular buffer: $\Delta'_{m}$, $ CIR_{m}$ and $\widehat{e_{m}}$. The reason for selecting the middle position in the buffer is to ensure that the averaged position, $p_{avg,m}$, is associated with the most representative data points in the buffer. By taking the middle position, you ensure that it is related to the same number of positions before and after. If you associate this average with a data point at the beginning or end of the set, it may not accurately represent the data points around that specific position. This is because the average is biased towards the data points on the side where there are more points.
Finally, to get an improved range estimate,  $p_{avg,m}$ is converted back to a range:$\Delta_{avg,m}$, by calculating the Euclidean distance with the anchor $a_n$. This value is our current best estimate of the range and is used in the reward function:
\begin{equation}
    R_{m} = \frac{1}{\lvert \Delta'_{m} - \Delta_{avg,m}\rvert}
\end{equation}
The goal of the reward function is to provide a quantitative measure of the success of an agent's actions in the environment. By shaping the reward function appropriately, we can guide the agent to exhibit the desired behavior, namely improved range accuracy. This reward function gives higher rewards the closer the corrected range of the RL agent is to the current best estimate of the range.
Updating the neural network at every time step with one sample would be very inefficient. Therefore, the network is updated on batches of data that are sampled from a replay memory containing experiences $(CIR_{m},\widehat{e_m},R_{m})$ generated during the execution of the algorithm. There are several methods to sample from this memory, for this problem we opted for random sampling instead of prioritized sampling as we do not want to overfit certain experiences or have a lack of diversity in the sampled experiences. An illustration of the complete proposed methodology is shown in Figure \ref{fig:completeoverview} and the pseudocode is given in Algorithm \ref{alg:ddpg}.

\begin{algorithm}
  \caption{Self-supervised RL for error correction}\label{alg:ddpg}
  \KwData{
    Initialize replay memory $D$\;
    Initialize circular smoothing buffer $C$ with length $N$\;
    Initialize actor network $\mu$ and critic network $Q$ function with random weights $\theta$ and $\phi$\;
    Initialize target actor $\dot{\mu}$ with weights $\dot{\theta}=0$\;
    Initialize target critic $\dot{Q}$ with random weights $\dot{\phi}$
  }
  \While{episode $<$ training episodes}{
    \While{episode not done}{
      Get current data $\widehat{\Delta_t}$, $CIR_t$, and $\widehat{e_t}$\;
      With probability $\epsilon$, select random correction $a_{t}$\;
      Otherwise, $a_{t} = \dot{e_t} = \dot{\mu}(\widehat{\Delta_t},\dot{\theta})$\;
      Correct range estimate $\dot{\Delta'}_t = \widehat{\Delta_t } - \dot{e_t}$\;
      Determine $p_{KF,t}$ = $\text{Kalman\textunderscore Filter}$ $(\dot{\Delta'}_t)$\;
      Add $p_{KF,t}$ to circular buffer $C$\;
      \If{C is full}{
        $p_{avg,m} = (\frac{1}{N} \sum_{i=t}^{t+N-1} x_i, \frac{1}{N} \sum_{i=t}^{t+N-1} y_i)$\;
        Convert $p_{avg,m}$ to $\Delta_{avg,m}$\;
        Calculate $R_{m} = \frac{1}{\lvert \Delta'_{m} - \Delta_{avg,m}\rvert}$\;
        Store experience $d_m = (CIR_{m},\widehat{e_t},R_{m})$ in $D$\;
      }
      \If{Every $K$ steps}{
        Sample a random minibatch $b$ from $D$\;
        \ForEach{$d_{j}$ in $b$}{
          $y_j = R_{j} + \gamma\dot{Q}(CIR_j,\widehat{e_j}\mid\dot{\phi})$\;
        }
        Update critic by minimizing the loss:
        $L=\frac{1}{B}\sum_{j} (y_j-Q(CIR_j,\widehat{e_j}\mid\phi))^2$\;
        Update actor using sampled policy gradient:
        $J = -\frac{1}{B} \sum_{j} Q(CIR_j,\widehat{e_j}\mid\phi)$\;
      }
      \If{Every $T$ steps}{
        Update target critic: $\dot{\phi} \leftarrow \tau \phi + (1-\tau)\dot{\phi}$\;
        \If{$\mu$ sufficiently trained}{
          Update target actor: $\dot{\theta} \leftarrow \tau \theta + (1-\tau)\dot{\theta}$\;
        }
      }
    }
    }
\end{algorithm}
 The network architecture of the actor and critic is given in Table \ref{tab:networkarchitectures}. The actor network is based on the CNN from \cite{fontaine2023transfer}, a state-of-the-art supervised model. Three convolutional layers are trained to extract local time series features from each CIR. The kernel of size 16 captures more large-scale patterns while the two layers with smaller kernels capture more fine-grained patterns in the data. The progression from 128 to 32 channels is designed to gradually reduce the dimensionality while preserving features. The only difference with \cite{fontaine2023transfer} is that the actor network ends with scaling the output of a dense layer with a Tanh activation function instead of no activation in the final layer. The Tanh output is between -1 and 1, which leads to the final output being scaled to -1000 and 1000 which is equal to the action space. Not changing the activation function for the actor network leads to training instability and does not result in a good-performing system. Changing the activation function in the supervised CNN also improves the supervised performance. The influence of this change on the supervised CNN is demonstrated in the evaluation.

The critic network starts similar to the actor network, but the information of the CIR is encapsulated in 4 latent features. These 4 latent features are concatenated with the action selected by the actor. This is then further processed to a final layer with a Tanh activation, meaning that the Q-value is between -1 and 1.

\begin{table*}[]
\centering
\caption{Actor and Critic Network Architectures}
\label{tab:networkarchitectures}
\begin{tabular}{lcclcc}
{\fontsize{10}{12}\selectfont \textbf{Actor network}} &  &  & {\fontsize{10}{12}\selectfont \textbf{Critic Network} }& \textbf{} & \textbf{} \\ \hline
\textbf{Layer} & \textbf{Activation} & \textbf{Output size} & \textbf{Layer} & \textbf{Activation} & \textbf{Output size} \\ \hline
\rowcolor[HTML]{E8F4FF} Input (State) &  & (1,150) &  Input (State) &  & (1,150)  \\
\rowcolor[HTML]{E8F4FF}  Conv1D(128,16) & ReLU & (128,150) &   Conv1D(128,16) & ReLU & (128,150)  \\
\rowcolor[HTML]{E8F4FF} Maxpool(2) &  & (128,75) &  Maxpool(2) &  & (128,75)  \\
\rowcolor[HTML]{E8F4FF}  Conv1D(64,8) & ReLU & (64, 75) &  Conv1D(64,8) & ReLU & (64, 75)\\
\rowcolor[HTML]{E8F4FF}  Conv1D(32,2) & ReLU & (32,75) &  Conv2D(32,2) & ReLU & (32,75) \\
\rowcolor[HTML]{E8F4FF} BatchNorm &  & (32,75) & BatchNorm &  & (32,75)\\
\rowcolor[HTML]{E8F4FF} Dropout 25\% &  & (32,75) & Dropout 25\% &  & (32,75)\\
\rowcolor[HTML]{E8F4FF} Flatten &  & 2400 &Flatten &  & 2400\\
\rowcolor[HTML]{E8F4FF} Dense & ReLU & 150 &Dense & ReLU & 150 \\
\rowcolor[HTML]{E8F4FF} BatchNorm &  & 150 &BatchNorm &  & 150\\
\rowcolor[HTML]{E8F4FF} Dropout 20\% &  & 150 & Dropout 20\% &  & 150\\
\rowcolor[HTML]{E8F4FF} Dense & ReLU & 100 &Dense & ReLU & 100\\
\rowcolor[HTML]{E8F4FF} Dropout 20\% &  & 100 &Dropout 20\% &  & 100 \\
\rowcolor[HTML]{E8F4FF} Dense & ReLU & 50 &Dense & ReLU & 50 \\
\rowcolor[HTML]{E8F4FF} Dropout 10\% &  & 50 &Dropout 10\% &  & 50 \\
\rowcolor[HTML]{E8F4FF} Dense & Sigmoid & 25 &Dense & Sigmoid & 25\\ \hline
\rowcolor[HTML]{E8E0FF} Dense & Tanh & 1 & Dense & ReLU & 4 \\
\rowcolor[HTML]{E8E0FF} Output Scaling (x1000) &  & 1 & Concat (add action) &  & 5 \\
 &  &  & \cellcolor[HTML]{E8E0FF}Dense & \cellcolor[HTML]{E8E0FF}ReLU & \cellcolor[HTML]{E8E0FF}8 \\
 &  &  & \cellcolor[HTML]{E8E0FF}Dense & \cellcolor[HTML]{E8E0FF}ReLU & \cellcolor[HTML]{E8E0FF}16 \\
 &  &  & \cellcolor[HTML]{E8E0FF}Dense & \cellcolor[HTML]{E8E0FF}ReLU & \cellcolor[HTML]{E8E0FF}8 \\
 &  &  &\cellcolor[HTML]{E8E0FF}Dense & \cellcolor[HTML]{E8E0FF}Tanh & \cellcolor[HTML]{E8E0FF}1
\end{tabular}
\end{table*}

\section{Results and analysis}
\label{sec:results}
\subsection{Baselines and metrics}
For performance evaluation, we will use two evaluation metrics: (1) the mean absolute error (MAE) as it encapsulates the performance in a single value and (2) box plots to provide a clear and concise way to see the spread (variability) of ranging errors and thus a more general overview of the performance while also highlighting central tendencies.
To evaluate our proposed method, we compare our results against two baselines: the first baseline is the uncorrected UWB performance, and the second is the state-of-the-art supervised CNN method \cite{fontaine2023transfer} trained on the fully labeled dataset. The results of the supervised CNN are not directly adopted from the paper itself, but the developed model has been retrained on the dataset of this research.
The range error results of NLOS samples will be shown and discussed separately because of the reduced signal clarity in NLOS situations. For NLOS, the signal propagation between transmitter and receiver is more complex due to the attenuated first path signal power, leading to a more complicated relationship between CIR and error correction. The NLOS situations are the most vital for error correction, as they are prone to the largest ranging errors. Showing the performance in NLOS situations separately provides insight into how the baselines and our proposed RL algorithm perform in the most challenging conditions.
\subsection{Training}
The RL algorithm was trained for 1000 episodes with $\gamma~=~0.5$, $\tau_{critic}~=\tau_{actor}~=~0.01$, $\alpha_{actor}~=~5e^{-5}$ and $\alpha_{critic}~=~5e^{-4}$. The patience of the learning rate schedulers was set to 150 episodes.  The batch size is 50 and the buffer size $N$ is set to 31 as this was shown to be a good trade-off. To evaluate the performance of our machine learning models, we split the dataset into training and testing sets. Specifically, we used an 80/20 split, where 80\% of the data was allocated for training the models, and the remaining 20\% was reserved for validation. Typically, RL does not require a distinct training and validation split, as the performance is usually evaluated based on the agent's interaction with the environment. However, since we are working with a predefined dataset and aim to make fair comparisons with supervised learning methods, we implemented this split.

The validation data is still used to maintain consistent trajectories and averaging. However, it is never incorporated into the replay buffer. This means that the actor and critic networks have not trained on the validation CIRs, ensuring that the RL algorithm has not been exposed to the CIR-correction combinations shown in the evaluations. This approach allows us to assess the generalization capabilities of our models. All the following evaluations use the first dataset that is described in section \ref{sec:dataset} except in subsection \ref{sec:adaptivity} where both are used.
\begin{figure}[ht]
    \centering
    \includegraphics[width=0.95\linewidth]{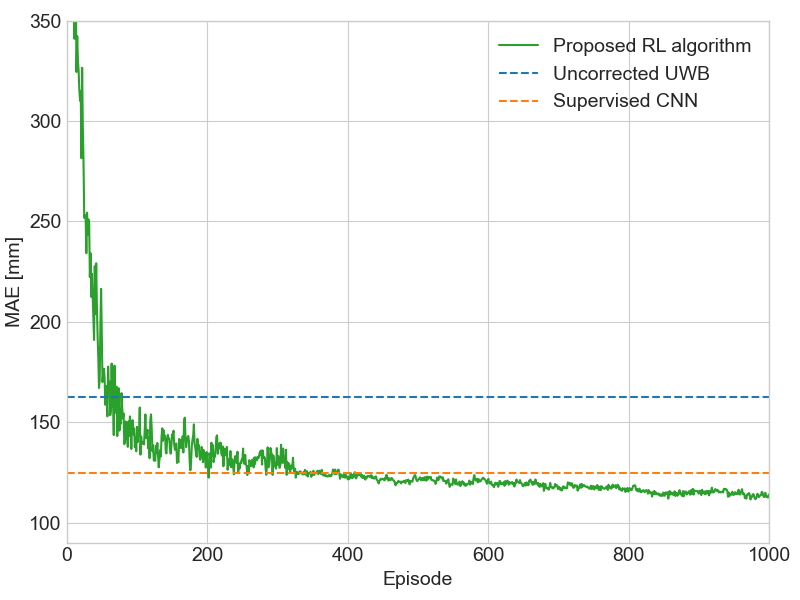}
    \caption{Performance comparison of our proposed RL algorithm during training with uncorrected UWB ranging and a supervised CNN approach in terms of MAE. The figure shows that our proposed algorithm quickly improves the ranging performance compared to uncorrected UWB ranging, and later surpasses the supervised CNN performance.}
    \label{fig:maetraining}
\end{figure}

Figure \ref{fig:maetraining} illustrates the learning curve of the algorithm. During the first 100 episodes, there is a steep decrease in MAE and thus a quickly improving performance. Between episodes 100 and 350, the decrease starts slowing down, which leads the scheduler to reduce the learning rates. The reduced learning rate is visible in Figure \ref{fig:maetraining} from episode 350 onwards. This early learning phase is primarily shaped by the exploration-exploitation trade-off, the exploration is decaying exponentially. Lower and faster decaying exploration would cause an even more steep decrease in MAE, but could come at the cost of worse final performance as the algorithm explores fewer possibilities. Higher and slower declining exploration would come at the cost of slower convergence and more training episodes needed.
In the figure, the reduced fluctuations in performance, from episode 350 onward, show the reduced learning rate.
At the end of the training, the MAE of the RL algorithm is distinctly lower than the uncorrected UWB ranging and the supervised CNN approach.
\begin{figure*}[ht]
    \centering
    \includegraphics[width=0.7\linewidth]{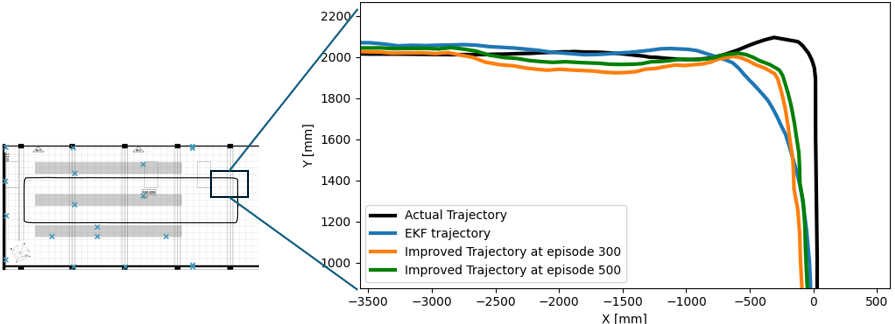}
    \caption{Trajectory comparison of the original EKF (with no RL correction) with the improved trajectories during training after 300 and 500 episodes}
    \label{fig:trajectory}
\end{figure*}
Figure \ref{fig:trajectory} illustrates the iterative improvement by the RL algorithm during training. The green curve represents the smoothed EKF trajectory without any RL correction, this is the data used to calculate the reward before the target actor is updated. Once the target actor starts getting updates from episode 350 onwards, the data used to calculate the rewards starts improving. The orange trajectory is used for reward calculation at episode 300 and the green trajectory is the improved trajectory at episode 500. This visually illustrates the iterative improvement of the algorithm.
\subsection{Evaluation}
\begin{figure*}[]
\centering
\begin{subfigure}[b]{0.40\textwidth}
\includegraphics[width=0.95\textwidth]{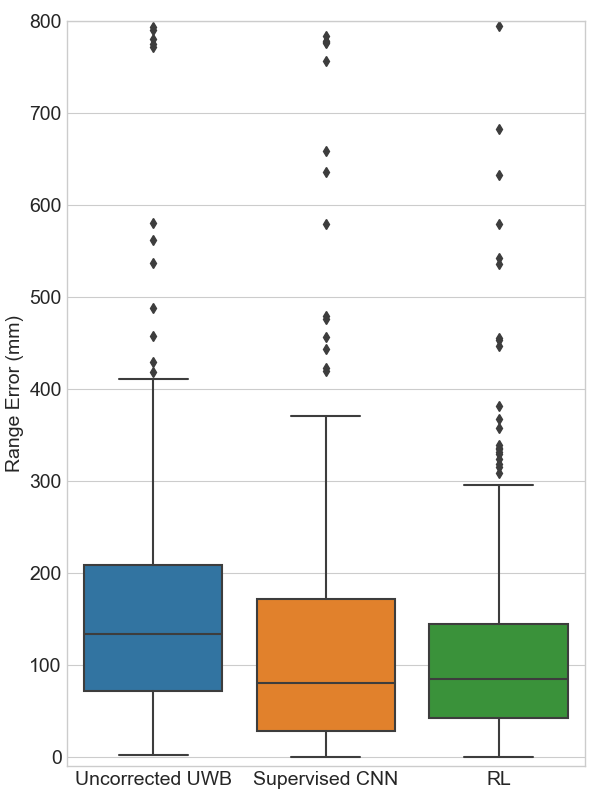}
\caption{All samples}
\label{fig:boxplotall} 
\end{subfigure}
\quad 
\begin{subfigure}[b]{0.40\textwidth}
\includegraphics[width=0.95\textwidth]{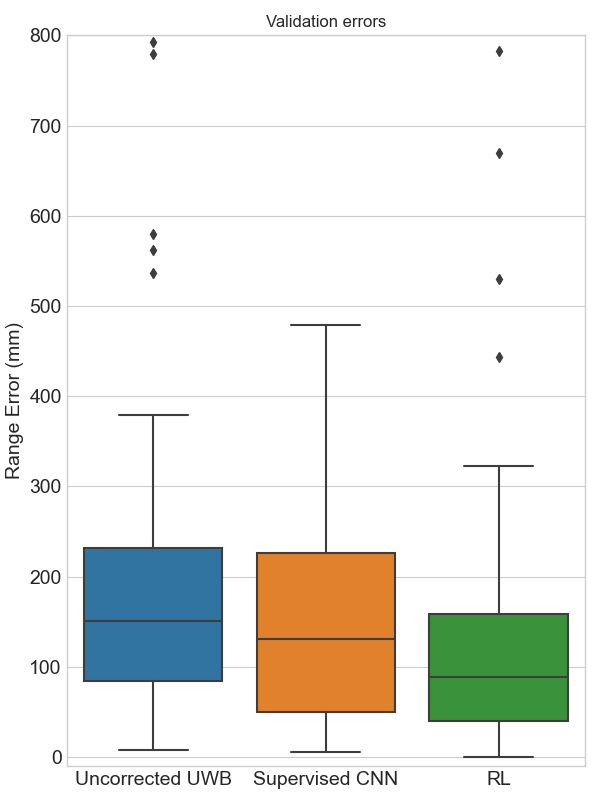}
\caption{Only NLOS samples}
\label{fig:boxplotsnlos}
\end{subfigure}
\caption{The ranging errors of uncorrected UWB, the supervised CNN, and our proposed RL algorithm during evaluation for (a) all samples and (b) only in NLOS samples. The figures show that our proposed self-supervised RL algorithm performs comparable or better than a supervised CNN approach.}
\label{fig:boxplots}
\end{figure*}

In Figure \ref{fig:boxplots}, the box plots show that the proposed RL algorithm performs better than the supervised CNN approach and significantly better than the uncorrected UWB.  The median range error of the RL algorithm is lower than the other two methods for the NLOS box plots but slightly higher than the median of the supervised CNN when comparing all evaluation samples. However, the interquartile range (IQR) is lower and thus more tightly clustered around the median. This indicates that the RL algorithm is more robust in its performance than the other two methods. Figure \ref{fig:boxplotsnlos} highlights the performance of the proposed RL algorithm in NLOS situations, with a lower median error rate and smaller interquartile range, this again indicates a more consistent performance in the more difficult situations. However, it has visibly more outliers than the supervised CNN approach.
Table \ref{tab:maeoverview} tells a similar story. The proposed RL approach reduces MAE by 31.6\% compared to uncorrected UWB and equals the supervised CNN from \cite{fontaine2023transfer}. The separate NLOS results emphasize the increased performance even more, as the proposed algorithm decreases the MAE by 34.8\% compared to uncorrected UWB and by 22.8\% compared to the supervised CNN.  Table \ref{tab:maeoverview} also contains the results of the optimized supervised CNN that uses the Tanh activation with output scaling instead of the linear layer. These results show that the improved accuracy compared to the supervised approach, visible in figure \ref{fig:maetraining} is mainly due to Tanh activation and scaling.

\begin{table}[ht]
\centering
\caption{Quantitative results of the baselines, the optimized version of the supervised CNN, and the proposed algorithm}
\label{tab:maeoverview}
\begin{tabular}{cccc}
\hline
Method & MAE (mm) & MAE (mm) \\ 
       & All Samples & NLOS \\ \hline
Uncorrected UWB & 162 & 194 \\
Supervised CNN \cite{fontaine2023transfer} & 124 & 153 \\
Optimized Supervised CNN & 112 & 137 \\
\textbf{Proposed RL Algorithm} & 112 & 128 \\ \hline
\end{tabular}
\end{table}

\subsection{Impact of changing environment}
\label{sec:adaptivity}
\begin{figure*}
    \centering
    \includegraphics[width=0.9\linewidth]{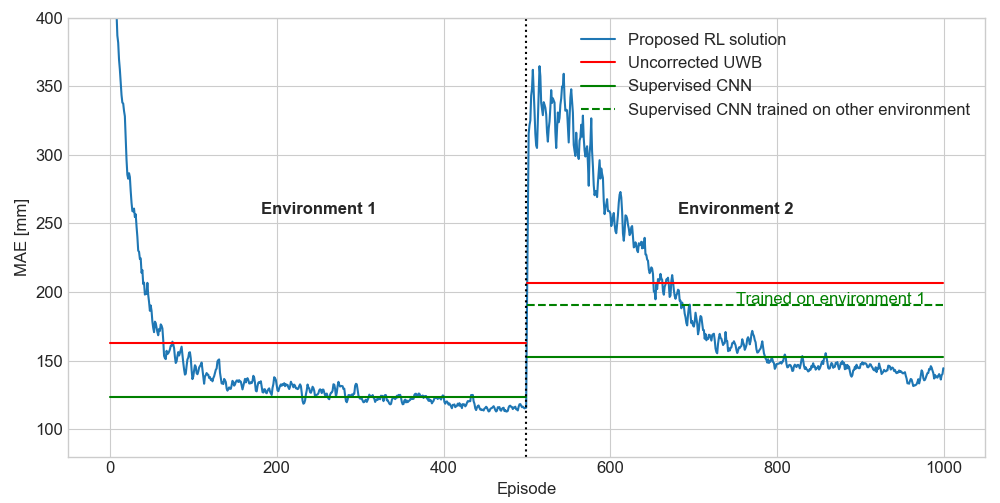}
    \caption{MAE comparison of our approach with a state-of-the-art CNN for error correction \cite{fontaine2023transfer} for a changing environment. The proposed RL algorithm is compared with uncorrected UWB and the supervised CNN trained on the current environment and the other environment. The figure shows the generalization problem of the supervised CNN and that the proposed RL algorithm can adapt itself to a changing environment.}
    \label{fig:adaptivity}
\end{figure*}
The same environment can change over time. To evaluate the performance of our algorithm when there are sudden changes in the environment, a new dataset was collected in the same warehouse 6 months later, as discussed in section \ref{sec:dataset}. Figure \ref{fig:adaptivity} shows the MAE during training. First, the proposed algorithm is trained on "environment 1", which is the original dataset used in the previous evaluation, between episodes 0-500, the training is similar to Figure \ref{fig:maetraining}, except it is halted after 500 instead of 1000 episodes. After 500 episodes, the environment is switched to "Environment 2", which is more difficult. The learning rates of the RL algorithm are reset to the starting values and the exploration is increased. Our reason for resetting the learning rate was to accelerate the learning. The current learning rate was low due to convergence in "environment 1". In a new environment, we reset the learning rate because it is not converged anymore. This adjustment can be made during the system's deployment. When significant changes occur in the environment, such as a new layout or the addition of racks, products, or items, increasing the learning rate allows the system to adapt more rapidly. Conversely, when the environment remains relatively stable, a lower learning rate suffices to accommodate everyday changes.
The dotted line in Environment 2 is the accuracy of the supervised CNN trained in Environment 1 but executed in Environment 2. This is a realistic situation in which a trained model is executed in an environment. These results show that traditional CNNs start to degrade over time when normal changes are made to the environment. As such, even using a single environment, we show that adapting the neural network is necessary for real-world usability. Our proposed RL algorithm provides this.
At first, the RL algorithm leads to worse performance, due to the exploration, but quickly adapts to the environment and surpasses the supervised model trained on the first dataset and later also the supervised CNN trained on the new dataset. This result displays the adaptivity of the RL algorithm compared to the supervised CNN approach, without needing to label a dataset it can adapt to changing environments and continuously leads to improved ranging performance. The supervised CNN approach requires a new labeled dataset in a changed environment, while our approach does not.
\subsection{Impact of robot trajectory}
\begin{figure}
    \centering
    \includegraphics[width=0.95\linewidth]{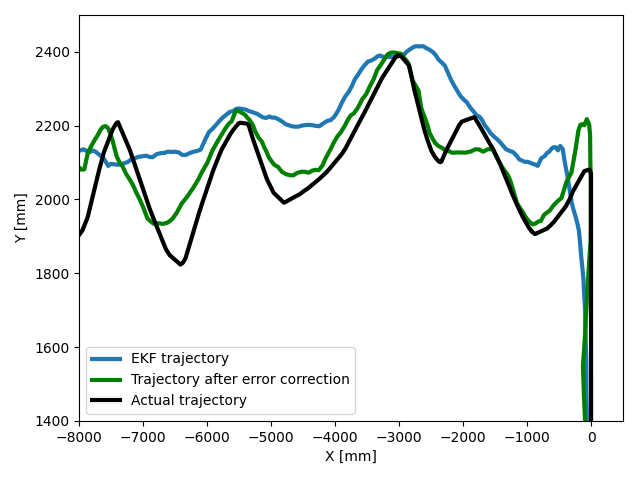}
    \caption{Once trained, our approach can also correct individual positions or unpredictable paths. The figure shows the evaluation of our approach for a trajectory with unpredictable paths, showing the original unpredictable path (in black - ground truth obtained using a MoCAP system), the corrected path using a traditional EKF approach (in blue), and the improved trajectory using our RL correction (in green).}
    \label{fig:random_traj}
\end{figure}

In training our neural network, we assumed the occasional presence of people or vehicles following predictable trajectories. However, after training on these predictable patterns, our neural network can accurately adjust to both individual positions and unpredictable trajectories. This section assesses the model's performance on a more unpredictable path to demonstrate that its learned behavior generalizes to other movement patterns. The data gathered over this unpredictable path, shown in Figure \ref{fig:random_traj} in black, consists of 300 UWB ranges with a MAE of 173 mm, after correction (using the trained actor model) the MAE is reduced to 128 mm. This improvement is in line with the previously reported improvements showing that our trained model also performs on more random trajectories. This improved ranging performance leads to an improved trajectory visible in Figure \ref{fig:random_traj}. 

\subsection{Complexity analysis}
\subsubsection{Algorithmic complexity}
The proposed RL algorithm leverages deep neural networks to approximate the actor and critic functions. Therefore, it is important to analyze their complexities. The networks can be broken down into different components. First, convolutional layers that can be calculated as follows: $O(H*W*C_{in}*C_{out}*K_w*K_h)$ with $H*W$ the input size, $C_{in}$ the input channels, $C_{out}$ the output channels and $K_w$x$K_h$ the kernel size.
Following Table \ref{tab:networkarchitectures}, this results in:
\begin{itemize}
    \item Conv1: $O(1*150*128*16) = O(3.07*10^5)$
    \item Conv2: $O(75*128*64*8) = O(4.92 * 10^6)$
    \item Conv3: $O(75*64*32*2)= O(3.07*10^5)$
\end{itemize}
Showing that the complexity is dominated by the second convolutional layer.
The complexity of a linear layer is given by $O(nm)$ with n the number of input features and m the number of output features. The first dense layer will be the largest and given by $O(3.60*10^5)$. Following linear layers become less and less complex as input and output sizes decrease. 
\subsubsection{Time complexity}
In terms of time, we have two constraints: (1) time required for (automatic) data capture (2) and time required for training and inference.
\textbf{Time required for (automatic) data capture}: we need about 200 episodes to retrain the network from scratch in a new environment (visible in Figure \ref{fig:adaptivity}). Each episode is about 3000 samples, sampled at 50 Hz, this leads to about 1 minute of data capturing per episode.  This means that our approach requires about 3 hours and 20 minutes of data capturing if starting the system from scratch (e.g. after installation). However, while the system is operational, this data is collected automatically, without requiring human intervention. As such, our system can be used in environments where data is collected continuously and keeps adapting at run-time. In contrast, the supervised CNN first necessitates an entire manual measurement campaign to collect a dataset with ground truth, followed by the model training process. The RL approach eliminates the need for a ground truth dataset, encapsulating the entire process within the algorithm itself.
\textbf{The training and inference time} was measured on two different hardware platforms. Training one batch containing 50 samples, (both actor and critic network) takes 0.09 seconds on an NVIDIA GeForce GTX 1080 Ti GPU and 0.19 seconds on an Intel(R) Core (TM) i7 8700 CPU. Due to the sampling time of the system being 50Hz, the models can be trained faster than the incoming data rate of the new samples. The inference time of the actor network (providing the corrections) on an NVIDIA GeForce GTX 1080 Ti GPU, is 0.0025 seconds, while on an Intel(R) Core (TM) i7 8700 CPU.  Given that the UWB localization system operates at a sampling rate of 50 Hz, a new range and CIR are received every 0.02 seconds. The way our proposed approach works is that each incoming sample first needs a correction from the anchor and then it is trained in batches of samples based on the replay buffer.  Generating correction for 50 samples thus takes 0.125 seconds on the GPU and 0.18 seconds on the CPU. Training on this batch takes an additional 0.09 or 0.19 seconds respectively. Combined this leads to a processing time of 0.18 and 0.37 seconds. Processing one second of incoming data takes less than one second. An overview of the timing aspects is given in Table \ref{tab:time-comparison}. In summary, our method offers a more flexible, efficient, and adaptable solution that can operate in real-time and continuously improve its performance without manual intervention, addressing the limitations of traditional CNN approaches in dynamic environments.
\begin{table*}[h]
\centering
\begin{tabular}{l|ll}
\hline
Time Aspect & \textbf{Our Method} & Supervised Method\\
\hline
Initial Data Collection & Automatic $\sim$3h 20min & Manual campaign (time varies) \\

Ground Truth Labeling & Not required & Required (Expensive, time-consuming) \\

Training Time (per batch) & 0.09s (GPU), 0.19s (CPU) & Not provided (typically longer) \\

Inference Time & 0.0025s (GPU), 0.0036s (CPU) & Not provided \\

Adaptation to New Environments & Continuous, automatic & Requires new data collection and training \\
\hline
\end{tabular}
\caption{Comparison of time aspects showing that our method offers a more flexible, efficient, and adaptable solution.}
\label{tab:time-comparison}
\end{table*}

\section{Future work}
\label{sec:future}
There are several avenues to further expand on this research. A first potential enhancement could be to make the EKF adaptive to the trajectory. Straight trajectories can have more smoothening, while corners need reduced smoothening. By modifying the smoothening and filtering process based on trajectory characteristics, the tracking system's accuracy and robustness can be further investigated.
A second area of improvement lies in the selection of points from the smoothed and filtered trajectory to the discrete data points for learning. Currently, the middle point in the buffer is associated with the average position. Future research could investigate the feasibility of defining a continuous trajectory and selecting the closest point. This adjustment could potentially lead to more responsive and accurate error correction processes.
Additionally, the system's capabilities could be expanded by integrating various sources of additional information. This includes exploring adding map data, reflections, CIR, and range data between anchor nodes (with known fixed positions). Furthermore, Inertial Measurement Unit (IMU) data could be added as input to the EKF to make it more robust and allow for better labels. 
Finally, apply this research to positioning systems using TDoA methods instead of TWR systems. TDoA, which relies on measuring the time delays of signals arriving at different nodes, is a widely used technique in wireless localization, and applying this research there would further improve the real-world practicality of more positioning systems.
\section{Conclusion}
\label{sec:conclusion}
In this work, we propose a novel self-supervised deep reinforcement learning approach for Ultra-Wideband ranging error correction that does not require ground truth data. This is significant because collecting large labeled datasets for model training is impractical for real-world indoor positioning system deployment. The methodology is based on the assumption that there are occasional movements of people or vehicles in the environment, following sufficiently predictable trajectories. Experiments on real-world measurements demonstrate our approach achieves comparable or improved ranging accuracy compared to a state-of-the-art CNN approach for error correction. Specifically, our method reduces errors by up to 31.6\% compared to uncorrected UWB in challenging situations without any data labeling. Additionally, the reinforcement learning agent can quickly adapt to changing environments. This makes our self-supervised framework highly practical for use in real indoor scenarios, as it removes the dependency on time-consuming and costly ground truth collection efforts. In summary, by not relying on labeled data, our approach paves the way for more scalable and generalized Ultra-Wideband error mitigation solutions using deep reinforcement learning that can be easily deployed in various indoor spaces.

\ifCLASSOPTIONcaptionsoff
  \newpage
\fi



%
\bibliographystyle{IEEEtran}
\bibliography{sample-base}

\begin{thebibliography}{10}
\providecommand{\url}[1]{#1}
\csname url@samestyle\endcsname
\providecommand{\newblock}{\relax}
\providecommand{\bibinfo}[2]{#2}
\providecommand{\BIBentrySTDinterwordspacing}{\spaceskip=0pt\relax}
\providecommand{\BIBentryALTinterwordstretchfactor}{4}
\providecommand{\BIBentryALTinterwordspacing}{\spaceskip=\fontdimen2\font plus
\BIBentryALTinterwordstretchfactor\fontdimen3\font minus \fontdimen4\font\relax}
\providecommand{\BIBforeignlanguage}[2]{{%
\expandafter\ifx\csname l@#1\endcsname\relax
\typeout{** WARNING: IEEEtran.bst: No hyphenation pattern has been}%
\typeout{** loaded for the language `#1'. Using the pattern for}%
\typeout{** the default language instead.}%
\else
\language=\csname l@#1\endcsname
\fi
#2}}
\providecommand{\BIBdecl}{\relax}
\BIBdecl

\bibitem{bazo2021survey}
R.~Bazo, C.~A. da~Costa, L.~A. Seewald, L.~G. da~Silveira, R.~S. Antunes, R.~d.~R. Righi, and V.~F. Rodrigues, ``A survey about real-time location systems in healthcare environments,'' \emph{Journal of Medical Systems}, vol.~45, pp. 1--13, 2021.

\bibitem{minne2019experimental}
K.~Minne, N.~Macoir, J.~Rossey, Q.~Van~den Brande, S.~Lemey, J.~Hoebeke, and E.~De~Poorter, ``Experimental evaluation of uwb indoor positioning for indoor track cycling,'' \emph{Sensors}, vol.~19, no.~9, p. 2041, 2019.

\bibitem{elsanhoury2022precision}
M.~Elsanhoury, P.~M{\"a}kel{\"a}, J.~Koljonen, P.~V{\"a}lisuo, A.~Shamsuzzoha, T.~Mantere, M.~Elmusrati, and H.~Kuusniemi, ``Precision positioning for smart logistics using ultra-wideband technology-based indoor navigation: A review,'' \emph{IEEE Access}, vol.~10, pp. 44\,413--44\,445, 2022.

\bibitem{huang2018location}
H.~Huang, G.~Gartner, J.~M. Krisp, M.~Raubal, and N.~Van~de Weghe, ``Location based services: ongoing evolution and research agenda,'' \emph{Journal of Location Based Services}, vol.~12, no.~2, pp. 63--93, 2018.

\bibitem{coppens2022overview}
D.~Coppens, A.~Shahid, S.~Lemey, B.~Van~Herbruggen, C.~Marshall, and E.~De~Poorter, ``An overview of uwb standards and organizations (ieee 802.15. 4, fira, apple): Interoperability aspects and future research directions,'' \emph{IEEE Access}, vol.~10, pp. 70\,219--70\,241, 2022.

\bibitem{alarifi2016ultra}
A.~Alarifi, A.~Al-Salman, M.~Alsaleh, A.~Alnafessah, S.~Al-Hadhrami, M.~A. Al-Ammar, and H.~S. Al-Khalifa, ``Ultra wideband indoor positioning technologies: Analysis and recent advances,'' \emph{Sensors}, vol.~16, no.~5, p. 707, 2016.

\bibitem{wymeersch2012machine}
H.~Wymeersch, S.~Maran{\`o}, W.~M. Gifford, and M.~Z. Win, ``A machine learning approach to ranging error mitigation for uwb localization,'' \emph{IEEE transactions on communications}, vol.~60, no.~6, pp. 1719--1728, 2012.

\bibitem{gifford2020impact}
W.~M. Gifford, D.~Dardari, and M.~Z. Win, ``The impact of multipath information on time-of-arrival estimation,'' \emph{IEEE Transactions on Signal Processing}, vol.~70, pp. 31--46, 2020.

\bibitem{denis2003impact}
B.~Denis, J.~Keignart, and N.~Daniele, ``Impact of nlos propagation upon ranging precision in uwb systems,'' in \emph{IEEE conference on Ultra Wideband Systems and Technologies, 2003}.\hskip 1em plus 0.5em minus 0.4em\relax IEEE, 2003, pp. 379--383.

\bibitem{mao2018probabilistic}
C.~Mao, K.~Lin, T.~Yu, and Y.~Shen, ``A probabilistic learning approach to uwb ranging error mitigation,'' in \emph{2018 IEEE Global Communications Conference (GLOBECOM)}.\hskip 1em plus 0.5em minus 0.4em\relax IEEE, 2018, pp. 1--6.

\bibitem{jaron}
J.~Fontaine, M.~Ridolfi, B.~Van~Herbruggen, A.~Shahid, and E.~De~Poorter, ``Edge inference for uwb ranging error correction using autoencoders,'' \emph{IEEE Access}, vol.~8, pp. 139\,143--139\,155, 2020.

\bibitem{li2023variational}
Y.~Li, S.~Mazuelas, and Y.~Shen, ``A variational learning approach for concurrent distance estimation and environmental identification,'' \emph{IEEE Transactions on Wireless Communications}, 2023.

\bibitem{li2021semi}
------, ``A semi-supervised learning approach for ranging error mitigation based on uwb waveform,'' in \emph{MILCOM 2021-2021 IEEE Military Communications Conference (MILCOM)}.\hskip 1em plus 0.5em minus 0.4em\relax IEEE, 2021, pp. 533--537.

\bibitem{fontaine2023transfer}
J.~Fontaine, F.~Che, A.~Shahid, B.~Van~Herbruggen, Q.~Z. Ahmed, W.~B. Abbas, and E.~De~Poorter, ``Transfer learning for uwb error correction and (n) los classification in multiple environments,'' \emph{IEEE Internet of Things Journal}, 2023.

\bibitem{li2023unsupervised}
Z.~Li, K.~Hu, T.~Wang, S.~Cui, and Y.~Shen, ``An unsupervised transfer learning method for uwb ranging error mitigation,'' \emph{IEEE Communications Letters}, 2023.

\bibitem{yangselfsupervised}
B.~Yang, J.~Li, Z.~Shao, and H.~Zhang, ``Self-supervised deep location and ranging error correction for uwb localization,'' \emph{IEEE Sensors Journal}, vol.~23, no.~9, pp. 9549--9559, 2023.

\bibitem{iiot}
\BIBentryALTinterwordspacing
 [Online]. Available: \url{https://www.ugent.be/ea/idlab/en/research/research-infrastructure/industrial-iot-lab.htm/}
\BIBentrySTDinterwordspacing

\bibitem{wipos}
B.~Van~Herbruggen, B.~Jooris, J.~Rossey, M.~Ridolfi, N.~Macoir, Q.~Van~den Brande, S.~Lemey, and E.~De~Poorter, ``Wi-pos: A low-cost, open source ultra-wideband (uwb) hardware platform with long range sub-ghz backbone,'' \emph{Sensors}, vol.~19, no.~7, p. 1548, 2019.

\bibitem{ADSTWR}
Y.~Jiang and V.~C. Leung, ``An asymmetric double sided two-way ranging for crystal offset,'' in \emph{2007 International Symposium on Signals, Systems and Electronics}, 2007, pp. 525--528.

\bibitem{DW1000}
``Decawave - dw1000 ic,'' {Accessed} March, 2024. [Online]. Available: https://www.decawave.com/product/dw1000-radio-ic/.

\bibitem{pytorch}
\BIBentryALTinterwordspacing
A.~Paszke, S.~Gross, F.~Massa, A.~Lerer, J.~Bradbury, G.~Chanan, T.~Killeen, Z.~Lin, N.~Gimelshein, L.~Antiga, A.~Desmaison, A.~Kopf, E.~Yang, Z.~DeVito, M.~Raison, A.~Tejani, S.~Chilamkurthy, B.~Steiner, L.~Fang, J.~Bai, and S.~Chintala, ``{PyTorch: An Imperative Style, High-Performance Deep Learning Library},'' in \emph{Advances in Neural Information Processing Systems 32}.\hskip 1em plus 0.5em minus 0.4em\relax Curran Associates, Inc., 2019, pp. 8024--8035. [Online]. Available: \url{http://papers.neurips.cc/paper/9015-pytorch-an-imperative-style-high-performance-deep-learning-library.pdf}
\BIBentrySTDinterwordspacing

\bibitem{kalmanfilter}
G.~Mao, S.~Drake, and B.~D. Anderson, ``Design of an extended kalman filter for uav localization,'' in \emph{2007 Information, Decision and Control}.\hskip 1em plus 0.5em minus 0.4em\relax IEEE, 2007, pp. 224--229.

\end{thebibliography}
%

\end{document}